\documentclass[12pt,A4paper]{article}
\usepackage[T1]{fontenc}
\usepackage[utf8]{inputenc}
\usepackage[english]{babel}
\usepackage{fancyhdr}
\usepackage{amsmath}
\usepackage{amssymb}
\usepackage{amsfonts} 
\usepackage{graphicx}
\usepackage{anysize}
\usepackage{cancel}
\usepackage{psfrag}
\usepackage{secdot}
\usepackage{sectsty}
\usepackage{cite}
\sectionfont{\fontsize{14}{15}\selectfont}
\subsectionfont{\fontsize{12}{15}\selectfont}

\begin{document}
\marginsize{1.5cm}{1cm}{1cm}{1cm}

\begin{center}
\mbox{\Large\textbf{Ramanujan's oscillator}}\\
\vspace{0.2cm}	
Gergely Nyitray\\
\textit{gergely.nyitray@mik.pte.hu}\\
\mbox{\small Department of Automation, Faculty of Engineering and Information Technology},\\ 
\mbox{\small University of Pécs, Boszorkány street 2, Pécs, Hungary.}
\end{center}

\section*{Abstract}

We aim to show that the dimensionless unit angular frequency of a certain mechanical oscillator is realized by a condition involving the Hardy-Ramanujan number 1729. This type of coupled oscillator can be called a Ramanujan's oscillator.

\section*{Introduction} 

\begin{figure}[h!]
 \centering
 \begin{psfrags}
 	\psfrag{k}[cc][cc]{$k$}
 	\psfrag{m}[cc][cc]{$m$}
 	\psfrag{m1}[cc][cc]{$m'$}
 	\psfrag{R}[cc][cc]{$R$}
    \psfrag{x1}[cc][cc]{$X$}
 	\psfrag{p1}[cc][cc]{$\varphi$}
 	\psfrag{P1}[cc][cc]{$\phi$}
 	\includegraphics[width=9cm]{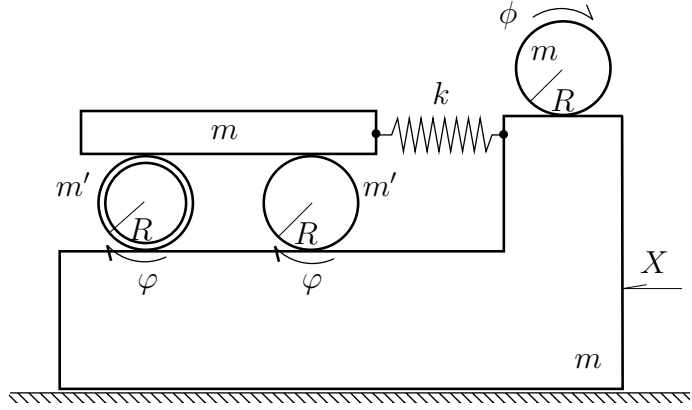}
	\caption{Coupled mechanical system.}
	\label{cms}
\end{psfrags}	
\end{figure}

Coupled mechanical systems can be successfully used as project work or sample task in higher level physics education because of their complexity and flexibility. These kinds of problems provide an opportunity to effectively apply various solution methods such as Euler-Lagrange and Hamilton's equations \cite{Goldstein,Marion}. Surprisingly, by their examination we can even get closer to number theory \cite{Stillwell}. We will show that the historically famous Hardy-Ramanujan number may appear during their solution. According to the story, the distinguished English mathematician G.~H.~Hardy once traveled to the sick Indian mathematician Srinivasa Ramanujan in taxi cab number 1729. Starting in 1914, Hardy was Ramanujan mentor and he was well aware Ramanujan's extraordinary mathematical skills. At the bedside of his sick colleague unable to come up with any comforting thoughts, Hardy began talking about the taxi number, remarking that it was quite boring. Ramanujan was immediately excited upon hearing this and explained to Hardy why this number was special. Ramanujan pointed out that this is the smallest number representable in two ways as a sum of two cubes\cite{Hardy,Brendt,Marcus}. It is given by $1729=1^3+12^3=9^3+10^3$. Ramanujan's unique abilities were very aptly described by J.~E.~Littlewood: ``Every positive integer is one of Ramanujan's personal friends'' \cite{Kanigel}. After the historical introduction, we would like to show that this number also arises when examining a particular mechanical oscillator, shown in Fig.~\ref{cms}: Two cylindrical rigid bodies are placed on the horizontal surface of a large block of mass $m$. The left one is a hoop and the other one is a cylinder. Both rigid bodies have the same mass $m'$ and radius $R$. On top of these bodies, a slab of mass $m$ is placed, which is fixed to the large block by an ideal spring with elastic constant $k$. Another cylinder of mass $m$ and radius $R$ is also placed on the higher horizontal surface of the large block. The system is initially at rest and the spring is not deformed. At a given moment, the two fixed points of the spring are slightly compressed horizontally and then released. It is assumed that during the motion neither the rolling bodies nor the slab slip, and the large block slides on the ground without friction. Our goal is to describe the motion of this oscillator.

\section{What is the natural frequency of the system?}

Under the conditions described above, the system will oscillate in normal mode, therefore let us calculate its angular frequency $\Omega$ if all parts of the system have the same mass ($m=m'$). Among the different solutions methods, we now present the one based on the Euler-Lagrange equations. The system has three degrees of freedom, since we have three independent coordinates which are $\varphi$, $\phi$ and $X$. If a cylindrical body of radius $R$ is rolling on a motionless surface without slipping, the instantaneous speed of its centre of mass (CM) is $R\dot{\varphi}$. The instantaneous speed of its highest point is twice as much ($2R\dot{\varphi}$). The orientations of the velocity vectors are parallel to the ground. If the surface is also moving in the opposite direction with instantaneous velocity $\dot{X}$, only velocities relative to the inertial reference frame (IRF) can be written into the Lagrangian as $\left(R\dot{\varphi}-\dot{X}\right)$. So the Lagrangian of the system takes the following form:
\begin{eqnarray}
	\mathcal{L}=\frac{1}{2}m\dot{X}^2+\frac{1}{2}m'\left(R\dot{\varphi}-\dot{X}\right)^2+\frac{1}{2}I_{\text{Hoop}}\,\dot{\varphi}^2+
	\frac{1}{2}m'\left(R\dot{\varphi}-\dot{X}\right)^2+\frac{1}{2}I'_{\text{Cyl}}\,\dot{\varphi}^2+\dots\nonumber\\ 
	\frac{1}{2}m\left(2R\dot{\varphi}-\dot{X}\right)^2+
	\frac{1}{2}m\left(R\dot{\phi}-\dot{X}\right)^2+\frac{1}{2}I_{\text{Cyl}}\,\dot{\phi}^2-\frac{1}{2}k\left(2R\varphi\right)^2,
\end{eqnarray}
where $I_{\text{Hoop}}$, $I'_{\text{Cyl}}$ and $I_{\text{Cyl}}$ denote the moment of inertia of the rolling bodies with respect to their symmetry axes:
\begin{equation*}
	I_{\text{Hoop}}=m'R^2,\quad I'_{\text{Cyl}}=\frac{1}{2}m'R^2\quad\text{and}\quad I_{\text{Cyl}}=\frac{1}{2}mR^2.
\end{equation*}
The total energy of a rolling body is obtained by adding the kinetic energy of the CM to the rotational energy. Since the deformation of the spring is independent of the overall position $X$ of the large block, so its value is determined solely by the instantaneous position of the highest points of the rolling rigid bodies (since the slab does not slip). So the deformation of the spring is $x=2R\varphi$ therefore the potential energy is $U=1/2kx^2=2kR^2\varphi^2$. 
After combining the same physical quantities into distinct terms the Lagrangian of the system takes the following form:
\begin{equation}
	\mathcal{L}=\frac{5}{2}m\dot{X}^2+\frac{3}{4}mR^2\dot{\phi}^2+\frac{15}{4}mR^2\dot{\varphi}^2-mR\dot{X}\dot{\phi}-
	4mR\dot{\varphi}\dot{X}-2kR^2{\varphi}^2. \label{la0}
\end{equation}
By introducing the Lagrange linear operator as 
\begin{equation}
	\Lambda_{i}=\frac{d}{dt}\left(\frac{\partial}{\partial \dot{q_i}}\right)-\frac{\partial}{\partial q_i}, 
\end{equation}
where $q_i$ denotes the independent coordinates ($X$, $\phi$, $\varphi$) so the relevant Euler-Lagrange equations can be written compactly:
\begin{subequations}
\begin{eqnarray}
	\Lambda_{X}\mathcal{L}&=&0\,\rightarrow\,\,	5\ddot{X}-R\ddot{\phi}-4R\ddot{\varphi}=0, \label{la1}\\
	\Lambda_{\phi}\mathcal{L}&=&0\,\rightarrow\,\,\frac{3}{2}R\ddot{\phi}-\ddot{X}=0, \label{la2}\\
	\Lambda_{\varphi}\mathcal{L}&=&0\,\rightarrow\,\,\frac{15}{2}mR\ddot{\varphi}-4m\ddot{X}+4kR\varphi=0. \label{la3}
\end{eqnarray}
\end{subequations}
Since during the normal mode oscillation all part of the system performs simple harmonic motion, therefore we are seeking its equation of motion in the usual form:
\begin{equation}
	\ddot{q}+\Omega^2q=0 \label{shm},
\end{equation}
where $\Omega=2\pi/T$. By eliminating $\ddot{\phi}$ from equations (\ref{la1}) and (\ref{la2}) gives
\begin{equation}
	\frac{15}{2}\ddot{X}-\ddot{X}=6R\ddot{\varphi}\,\rightarrow\,\ddot{X}=\frac{12}{13}R\ddot{\varphi}.
\end{equation}
Substituting  $\ddot{X}$ into equation (\ref{la3}), after simplifications we obtain the equation of motion
\begin{equation}
	\ddot{\varphi}+\left(\frac{104k}{99m}\right)\varphi=0 \label{shm2}.
\end{equation}
Comparing equation Eq.~(\ref{shm2}) with Eq.~(\ref{shm}), yields
\begin{equation}
 \Omega^2=\frac{104k}{99m}\approx1.0506\frac{k}{m}\,\rightarrow\,\frac{\Omega^2}{k/m}\approx 1. \label{approx}
\end{equation}
The following question arises: Under what conditions does the approximation in Eq.~(\ref{approx}) become an equality?

\section{The appearance of the number 1729}

Dimensionless unit angular frequency can be achieved by introducing mass differences between the elements that forming the system. More specifically the masses of the rigid bodies rolling in pairs, denoted by $m'$ and the masses $m$ of the other elements of the system are considered to be different. Then it is assumed that the unit angular frequency can be achieved by proper choice of the $m'/m$ ratio. Taking into account the differences in mass the Lagrangian of the re-parameterized system is slightly different from Eq.~(\ref{la0}):
\begin{eqnarray}
	\mathcal{L}=\frac{3m+2m'}{2}\dot{X}^2+\frac{3}{4}mR^2\dot{\phi}^2+\left(\frac{7}{4}m'+2m\right)R^2\dot{\varphi}^2\dots\nonumber\\
	-mR\dot{X}\dot{\phi}-2\left(m'+m\right)R\dot{\varphi}\dot{X}-2kR^2{\varphi}^2.
\end{eqnarray}
The relevant Euler-Lagrange equations are:
\begin{subequations}
	\begin{eqnarray}
		\Lambda_{X}\mathcal{L}&=&0\,\rightarrow\,\,	
		\left(3m+2m'\right)\ddot{X}-mR\ddot{\phi}-2\left(m+m'\right)R\ddot{\varphi}=0, \label{la12}\\
		\Lambda_{\phi}\mathcal{L}&=&0\,\rightarrow\,\,	
		\frac{3}{2}R\ddot{\phi}-\ddot{X}=0, \label{la22}\\
	    \Lambda_{\varphi}\mathcal{L}&=&0\,\rightarrow\,\,	
		\left(\frac{7}{2}m'+4m\right)R\ddot{\varphi}-2\left(m+m'\right)\ddot{X}+4kR\varphi=0. \label{la32}
	\end{eqnarray}
\end{subequations}
Substituting $\ddot{X}=3/2R\ddot{\phi}$ from Eq.~(\ref{la22}) into Eq.~(\ref{la12}) yields
\begin{subequations}
\begin{eqnarray}
	\ddot{X}&=&\frac{2\left(m'+m\right)}{2m'+2m+\frac{m}{3}}R\ddot{\varphi}\label{con1},\quad\text{and}\\
    \ddot{\phi}&=&\frac{4\left(m'+m\right)}{6m'+7m}\ddot{\varphi}\label{con2}.
\end{eqnarray}
\end{subequations}
After substituting $\ddot{X}$ from Eq.~(\ref{con1}) into Eq.~(\ref{la32}) the equation of motion takes the following form:
\begin{equation}
	\ddot{\varphi}+\left(\frac{8k}{7m'+8m-\frac{8\left(m'+m\right)^2}{\rule[-2mm]{0pt}{14pt}\frac{7m}{3}+2m'}}\right)\varphi=0.
\end{equation}
It is easy to see that if the first and third terms in the denominator of $\Omega^2$ is just cancel out each other, 
the dimensionless unit angular frequency has been realized, hence
\begin{eqnarray}
	7m'-\frac{8\left(m'+m\right)^2}{\rule[-2mm]{0pt}{14pt}\frac{7m}{3}+2m'}=0\,\rightarrow\,18{\varepsilon}^2+{\varepsilon}-24=0,
\end{eqnarray}
where $\varepsilon=m'/m$. The physically acceptable solution of this quadratic equation is:
\begin{eqnarray}
	\varepsilon=\frac{-1+\sqrt{1-4\cdot18\cdot(-24)}}{36}=\frac{\sqrt{1729}-1}{36}\approx 1.127256827.\label{dyn}
\end{eqnarray}
 Notice that the number 1729 in Eq.~(\ref{dyn}) not only appears, but interestingly, appears as a sum of two cubes, just as Ramanujan referred to it: $\sqrt{1-4\cdot18\cdot(-24)}=\sqrt{1^3+12^3}$.

\subsection{Dynamic Interpretation of number 1729}

Based on Eq.~(\ref{dyn}), the Hardy-Ramanujan number can also be defined by the $\varepsilon=m'/m$ ratio at which our specific  oscillator vibrates with unit angular frequency:
\begin{eqnarray}
	1729=\left(36\varepsilon+1\right)^2.
\end{eqnarray}

\section{Relationship between the natural frequencies}

The natural frequencies of the three modes can be easily determined.
The relationship between the independent variables are $\ddot{X}=c_1\ddot{\varphi}$, $\ddot{X}=c_2\ddot{\phi}$ and $\ddot{\varphi}=\left(c_2/c_1\right)\ddot{\phi}$, where the constants $c_1$ and $c_2$ can be identified based on Eqs.~(\ref{con1}) and (\ref{con2}).
Let us consider the equation of motion of the system ($\ddot{\varphi}+\Omega_{\varphi}^2\,\varphi=0$) and replace the variable $\varphi$ by the variable $X$ and $\phi$ taking into account that $\varphi=X/c_1$ and  $\varphi=\left(c_2/c_1\right)\phi$ 
\begin{subequations}
\begin{eqnarray}
\ddot{X}c_1^{-1}&=&\Omega_{\varphi}^2\,\left(Xc_1^{-1}\right)\,\rightarrow\,\ddot{X}=\Omega_{\varphi}^2\,X\,\rightarrow\,\Omega_{X}=
\Omega_{\varphi},\\
\ddot{\phi}\,c_2c_1^{-1}&=&\Omega_{\varphi}^2\,\left(\phi \,c_2c_1^{-1}\right)\,\rightarrow\,\ddot{\phi}=\Omega_{\varphi}^2\,\phi\,\,\rightarrow\,\Omega_{\phi}=\Omega_{\varphi}.
\end{eqnarray}
\end{subequations}
Hence all three modes have the same natural frequency:
\[
\Omega_X=\Omega_{\phi}=\Omega_{\varphi}.
\]

\section{Conclusion}

It has been shown that the dimensionless unit angular frequency of a certain type of coupled oscillator is realized by a condition that includes the Hardy-Ramanujan number. First, the natural frequency of the system was calculated by considering the mass of all elements to be the same. In this case the calculated dimensionless natural frequency squared turned out to be close to unity. We obtained exactly unit value by slightly increasing the masses of the rigid bodies rolling in pairs relative to the other elements of the system. Since this ratio can only be expressed in terms of the Hardy-Ramanujan number this gives opportunity for its dynamic interpretation. It is possible that there is a number theoretic reason for the appearance of the number 1729, although this remained hidden from the author for the moment.

\section{Conflict of Interest Disclosure statement}

The author has no conflicts to disclose.

\end{document}